\documentclass[aps,pre,showkeys,preprint]{revtex4-1}
\usepackage{lipsum}
\usepackage[T1]{fontenc}
\usepackage[utf8]{inputenc}
\usepackage{amssymb,bm}
\usepackage{amsmath,amsfonts,relsize}
\usepackage{graphicx,hyperref,xcolor}
\usepackage{lineno,setspace}
\usepackage{verbatim,xcolor}
\begin{document}
\title{Energy exchanges in a damped Langevin-like system with two thermal 
baths and an athermal reservoir}
%
%
%
\author{E. S. Nascimento}
 \email{edusantos18@esp.puc-rio.br}
 \affiliation{Dept. of Physics, PUC-Rio, Rua Marquês de São Vicente 225, 22451-900, Rio de Janeiro, Rio de Janeiro, Brazil}
\author{W. A. M. Morgado}
 \email{welles@puc-rio.br}
 \affiliation{Dept. of Physics, PUC-Rio, Rua Marquês de São Vicente 225, 22451-900, Rio de Janeiro, Rio de Janeiro, Brazil}
 \affiliation{National Institute of Science and Technology for Complex Systems, Brazil}
%
%
%
%
%
%
%
\begin{abstract}
We study a Langevin-like model which describes an inertial particle in 
a one-dimensional harmonic potential and subjected to two heat baths and one athermal environment. The thermal noises are white and Gaussian, and the temperatures of heat reservoirs are different. 
The athermal medium act through an external non-Gaussian noise of Poisson type. We calculate exactly the time-dependent cumulant-generating function of position and velocity of the particle, as well as an expression of this generating function for stationary states. 
We discuss the long-time behavior of first cumulants of the energy injected by the athermal reservoir and the heat exchanged with thermal baths. In particular, we find that the covariance of stochastic heat due to distinct thermal baths exhibits a complex dependence on properties of athermal noise.
\end{abstract}
\maketitle
%
%
%
%
%

%
%
\section{Introduction} \label{Intro}

Equilibrium states of macroscopic systems are usually driven to change through energy transfers in the 
form of heat and work with the external environment \cite{Callen1960}. According to thermodynamic 
formalism, the internal energy is a state function in thermodynamic equilibrium, but heat and work 
depend on the process in consideration: they are not state functions. Indeed, these two quantities 
were shown to be equivalent, in the seminal work of Joule~\cite{Joule1850}. However, for small-scale physical systems, heat and work are process-dependent quantities as well as time-dependent random variables 
\cite{Sekimoto2010,Sekimoto1998}, which exhibit intriguing properties because of the presence of inherent fluctuations \cite{BustamanteRitort2005,Seifert2012,Holubec_livro,Ciliberto2017}. 
A paradigmatic system used to study those fluctuating properties is a Brownian particle coupled to one or many thermal baths and under the action of conservative and external forces 
\cite{SpeckSeifert2007,GomezSolano2010,SahaJayannavar2009,MorgadoSoares-Pinto2010,BlickleBechinger2011,
FogedbyImparato2014,
ArgunVolpe2017,AlbayJun2021,HolubecMarathe2020}, where the distribution functions of heat and work can be determined in some cases \cite{GomezSolano2010,MorgadoSoares-Pinto2010,Sabhapandit2012,Ciliberto2017,Holubec_livro}. In fact, one may assume that the Brownian particle is confined in a harmonic potential \cite{SahaJayannavar2009,MorgadoSoares-Pinto2010,FogedbyImparato2011}, which simplifies the theoretical analysis of the problem, in addition to be implemented experimentally 
\cite{WangEvans2002,BlickleBechinger2011,ArgunVolpe2017}.

Due to the importance of fluctuating effects on the energetics of small scale systems, there is also an interest in the investigation of the statistics of heat and work in Brownian models with external stochastic forces \cite{Farago2002,Sabhapandit2012,PalSabhapandit2014,
GuptaSabhapandit2017,TouchetteCohen2007,BauleCohen2009,
MedeirosQueiros2015,Queiros2016}, such as external Gaussian \cite{Farago2002,Sabhapandit2012,PalSabhapandit2014,GuptaSabhapandit2017} and non-Gaussian noises \cite{TouchetteCohen2007,BauleCohen2009,
MedeirosQueiros2015,Queiros2016}. Experimental setups involving external noises have been proposed to study fluctuations on mesoscopic scales \cite{MestresRoldan2014,DinisRica2016,
Rossnagel2016,ParkPak2020,RoyGanapathy2021}. Probably, a simple model system driven by an 
external noise is a charged Brownian particle (charge $q$) in contact with a fluid, which plays the 
role of a thermal bath, that lies inside the plates of a capacitor large enough to contain the system and the fluid. The $x$-axis is the direction orthogonal to the plates; $\Delta \phi \left( t \right)$ is the 
instantaneous voltage difference between the capacitor plates that can evolve in a stochastic fashion. Let us consider that the voltage varies 
with a typical frequency  $\omega_0$ such as $\omega_0\, d \ll c$, where 
$c$ is the speed of light, and negligible electric effects on thermal bath substance. 
Thus, the energy of the particle due 
to the voltage difference is 
$
-x \,q\, \Delta \phi \left( t \right)/d,
$
where $d$ is the distance between the plates. 
The force experienced by the particle is simply $F \left( t \right)\equiv q\, \Delta \phi \left( t \right)/d$. 
The linear form of the time-dependent energetic coupling obtained is typical in linear response theory~\cite{Kubo1957}. 
In fact, the voltage $\Delta \phi(t)$ drives the averages of the physical quantities associated with the Brownian particle~\cite{hansenmcdonald,GrootMazur}. For the case of high-frequency components that are reasonably bounded, $\Delta \phi(t)$ can assume an arbitrary time-dependence. If the distance $d$ is small enough, the cut-off frequency $c/d$ tends to infinity. Then, we might suppose that the spectrum of $\Delta \phi \left( t \right)$ covers the range $(-\infty,\infty)$. For instance, in the experimental settings of the Paul trap at reference~\cite{Rossnagel2016}, where a single-ion heat engine is implemented, an externally generated electric noise field is used to shake the ion, imparting it with energy, analogous to a heat source. The electric shaking acts as a mechanism that ``heats up'' the system (ion). Also, external stochastic forces can be used in experiments involving colloidal particles in water and optically trapped by effective harmonic forces \cite{MestresRoldan2014,DinisRica2016}.  In addition, a confined Brownian particle with non-Gaussian noise is studied experimentally in the context of stochastic engines \cite{RoyGanapathy2021}.

In the present work, we explore the effects of a time-dependent force $F\left( t \right)$ on the energetic 
currents related to a Langevin-like system, where $F\left( t \right)$ is external and behaves approximately 
like a Poisson white noise \cite{MorgadoSoares2011,MorgadoQueiros2016}, which can be considered as a kind of 
athermal noise~\cite{Kanazawa_livro,KanazawaHayakawa2013,KanazawaHayakawa2015}. Since a Gaussian noise may 
be obtained from a Poisson noise \cite{van_Kampen-book}, we can investigate how Gaussian and non-Gaussian 
fluctuations induced by an external noise influence the energetics of Brownian-type systems. In order to consider a general non-equilibrium case, we assume the system is also affected by two heat reservoirs at different temperatures. The model describes a particle with non-zero mass in a one-dimensional harmonic potential and coupled to two thermal baths and an athermal reservoir. The cumulant-generating function of position and velocity of the particle is obtained exactly, which shows that the associated joint distribution is non-Gaussian due to the Poissonian driving force. We study the stationary properties 
of the first cumulants of heat and time-integrated power associated with external noise. 
In steady-state regime, the system experiences non-zero mean energetic currents due to the presence of thermal and non-thermal reservoirs, which characterize a non-equilibrium behavior. The fluctuations of these flows of energy are affected by the kind of external noise, i.e., if the stochastic force is Gaussian or Poissonian. Also, we find that the properties of the external noise play a role in the correlations between heat exchanged with distinct thermal baths.

We organize the paper as follows. In Sec. \ref{LangevinModel}, we define the system as a Langevin-like model. We determine the cumulant-generating function of position and velocity in Sec. \ref{ProDisXV}. We discuss the stochastic properties of energetic exchanges in Sec. \ref{StotchEner}. 
Finally, in Sec. \ref{Con} we present the conclusions. 

\section{Langevin-like dynamics} \label{LangevinModel}
\begin{figure}[h]
 \centering
 \includegraphics[scale=0.35]{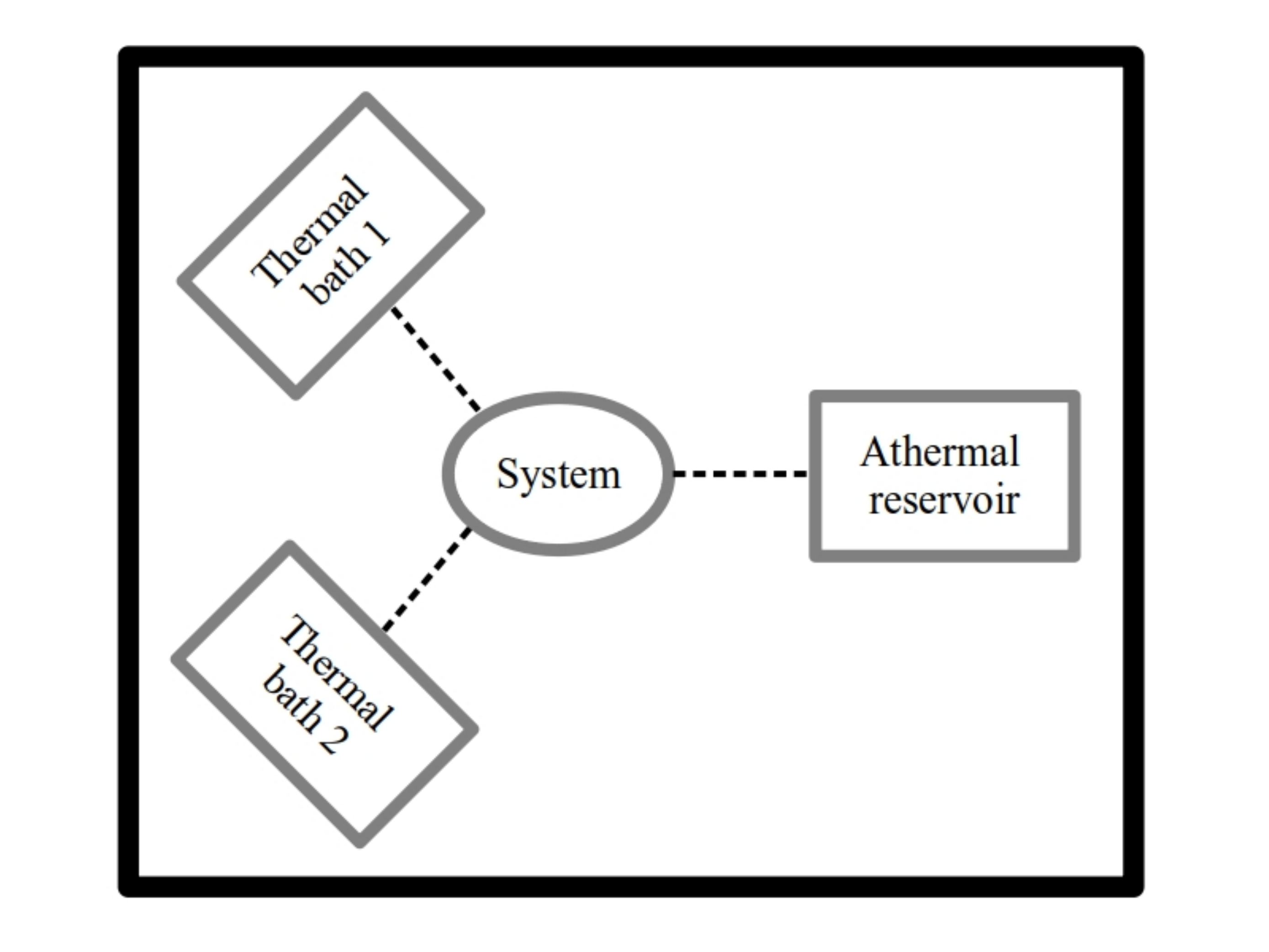}
 \caption{Schematic representation of a physical system simultaneously influenced by two thermal baths and an athermal reservoir. }
\label{Schematic}
\end{figure}

The model we are interested in consists of a damped harmonic oscillator with different types of stochastic forces. One can imagine that the confined inertial particle represents a system coupled to two thermal reservoirs and an athermal environment, see Fig. \ref{Schematic}. The system evolves in time according to the Langevin-like dynamics
\begin{equation} \label{LanSys1}
 \begin{split}
  m \dot{V}\left( t \right) &= - \left( \gamma_1 + \gamma_2 \right) V\left( t \right) - kX \\
  & \quad + \xi_1\left( t \right) + \xi_2\left( t \right) +  F\left( t \right), \\
  \dot{X}\left( t \right) &= V\left( t \right),
 \end{split}
\end{equation}
where $X$ is the position,  $V$ is the velocity, $m$ is the particle's mass, $\gamma_i$ is a friction coefficient ($i=1,2$), $k$ is the spring parameter, $\xi_i$ is the thermal noise related to heat reservoir $i$, 
and $F$ is the external noise that characterizes the non-thermal  reservoir. In this work, we assume the initial conditions $X(0)=0$ and $V(0)=0$, which is probably the simpler situation we can adopt here. However, in a more general case, it is important to bear in mind that the initial conditions may play a prominent role in the stochastic behavior of energy flows of the model \cite{LeePark2013}. We are also supposing that the noises $\xi_i$ and $F$ are independent.

The Langevin force $\xi_i$ is a white Gaussian noise with cumulants 
\begin{equation} \label{NoiseCumuGauss}
 \begin{split}
 \left<\left<  \xi_i(t)  \right>\right> &= 0, \\
 \left<\left< \xi_i(t_1)\xi_j(t_2) \right>\right> &= 2\gamma_i T_i \delta_{ij}\delta\left( t_1 - t_2 \right),
 \end{split}
\end{equation}
where 
$T_i$ is the effective temperature of thermal bath $i$ (the Boltzmann constant is assumed as $k_B\equiv 1$); $\delta_{ij}$ and $\delta\left( t_1 - t_2\right)$ are respectively Kronecker and Dirac delta functions. We adopt the notation of double angle brackets for cumulants and single brackets 
for moments.
The stochastic force $F$ is an external 
white noise with non-Gaussian properties \cite{van_Kampen-book}. For simplicity, we consider that $F$ is unbiased and symmetric.
The non-zero cumulants of the athermal noise are converging and given by 
\begin{equation} \label{NoiseCumuPoisson}
 \begin{split}
 \left<\left< F(t) \right>\right> &= 0, \\
 \left<\left< F(t_{1})F(t_{2}) \cdots F(t_{2n}) \right>\right> &= 
  C_{2n}\prod_{j=1}^{2n-1} \delta\left( t_{j +1} - t_{j}  \right). \\
 \end{split}
\end{equation}
The odd cumulants are zero by construction, yielding an infinite number of even cumulants~\cite{Marcinkiewicz1939}.
The positive quantity $C_{2n}$ presents dimension of impulse to the power of $2n$ over one unit time ($N^{2n}s^{2n-1}$). In practice, we consider that $F$ is an unbiased Poisson noise with rate of events $\lambda$ and impulses $J$ that follow a Gaussian distribution $\rho \left( J \right)$ with zero mean and standard deviation $b$ \cite{LuczkaHanggi1997,MorgadoQueiros2016}: 
\begin{equation} \label{ImpulseDist}
 \rho \left( J \right) = \frac{1}{\sqrt{2\pi b^2}} \exp\left( -\frac{J^2}{2b^2}
\right).\end{equation}
Then, one finds 
\begin{equation} \label{EqC2n}
 C_{2n} = \left( 2n-1\right)!! \lambda  b^{2n}.
\end{equation}
Observe that, although we are supposing an Poisson stochastic force,  it is possible to obtain an external Gaussian noise by taking the limit 
$\lambda \to \infty$, $b^2 \to 0$, with the product 
$ \lambda b^2 $ being constant \cite{van_Kampen-book,LuczkaHanggi1997,MorgadoQueiros2016}. This limiting case may be understood by writing $C_{2n}$ in terms of $C_2$ as $C_{2n}=\left(2n-1\right)!! \lambda^{1-n}C_2^{n}$, 
with $C_2 = \lambda b^2$ fixed. Thus, if $\lambda$ is very large and $n>1$, we have $C_{2n} \to 0$ and only $C_2$ is non-zero when external kicks act in a very frequent way. Therefore, the type of external noise $F$ considered allows us to investigate the effects of athermal Gaussian and non-Gaussian fluctuations on a Brownian-type model with two distinct heat baths.

\section{Cumulant-generating function of position and velocity} \label{ProDisXV}

The instantaneous position and velocity associated with the linear Langevin-like model \eqref{LanSys1} can be determined through the approach discussed in \cite{deOliveira2015,Zwanzig}, even though we have a Poissonian noise. We write \eqref{LanSys1} in a matrix form
\begin{equation} \label{MatrxiLanSys}
   \dot{Y}\left(t \right) = \mathbb{A} Y\left( t \right) + \Omega\left( t \right), \quad Y(0) = 0,
\end{equation}
where
\begin{equation}
  Y\left( t \right) = 
  \begin{pmatrix}
   X\left( t \right)\\
   V\left( t \right) 
  \end{pmatrix}, \quad 
   \mathbb{A} = \frac{1}{m}
  \begin{pmatrix}
   0 & m \\
   -k & -\gamma 
  \end{pmatrix}, \quad 
  \Omega\left( t \right) = 
  \begin{pmatrix}
   0 \\
   \omega\left( t\right)
  \end{pmatrix},
\end{equation}
with $\gamma = \gamma_1 + \gamma_2$ and $ \omega\left( t\right) = F\left(t\right) + \xi_1\left(t\right) + \xi_2\left(t\right)$. 
Notice that \eqref{MatrxiLanSys} resembles a non-homogeneous system of linear differential equations with constant coefficients. Since $\mathbb{A}$ is time-independent, the formal solution of \eqref{MatrxiLanSys} is
\begin{equation} \label{FormalSol}
 Y\left( t \right) = \int_{0}^{t} ds \exp\left[\left(t-s \right) \mathbb{A} \right] \Omega\left( t \right).
\end{equation}
We calculate the exponential of matrix $\mathbb{A}$ by using its eigenframe system:
\begin{equation} \label{ExpMatrix}
 \exp\left( s\mathbb{A} \right) = \mathbb{M}_{R}\exp\left( s\mathbb{D} \right) \mathbb{M}_{R}^{-1},
\end{equation}
where $\mathbb{D}$ is the diagonal form of $\mathbb{A}$, $\mathbb{M}_{R}$ is a matrix those columns are the right eigenvectors of $\mathbb{A}$, and $\mathbb{M}_{R}^{-1}$ is the inverse of $\mathbb{M}_{R}$. Suppose that $\nu$ is an eigenvalue of $\mathbb{A}$ with associated right eigenvector $R$. Then, after some calculation, one finds
\begin{equation} \label{Para1}
 \begin{split}
\nu_{\pm} = -\frac{\gamma}{2m} \left( 1 \pm i\sqrt{\frac{4km}{\gamma^2} -1} \right), \quad R_{\pm} = 
 \begin{pmatrix}
  1 \\
  \nu_{\pm}
 \end{pmatrix},
 \end{split}
\end{equation}
where $i = \sqrt{-1}$ is the imaginary unit.
Therefore, using \eqref{Para1} in \eqref{ExpMatrix}, we obtain
\begin{equation} \label{XeqVeq}
\begin{split}
 X(t) = \int_0^t ds\, A_{x}\left( t-s \right)\left[ F\left(s\right) + \xi\left(s\right) \right], \quad V(t) = \int_0^t ds\,A_{v} \left( t -s\right)\left[ F\left(s\right) + \xi\left(s\right) \right],
\end{split}
\end{equation}
where $\xi\left(s\right)=\xi_1 \left( s \right)+ \xi_2\left( s \right)$, and
\begin{equation} \label{AxAv}
 \begin{split}
 A_{x} \left( t \right) = \frac{ e^{\nu_{+} t } - e^{\nu_{-} t }}{m\left( \nu_+ - \nu_-\right)}, \quad A_{v} \left( t \right) = \frac{ \nu_+ e^{\nu_{+} t } - \nu_- e^{\nu_{-} t }}{m\left( \nu_+ - \nu_-\right)},
 \end{split}
\end{equation}
Observe that $\nu_{-}$ is the complex conjugate of $\nu_{+}$, and due to that $A_{x,v}$ are real-valued quantities. 

The formal solutions \eqref{XeqVeq} allow us to determine the time-dependent cumulants of $X$ and $V$. 
Indeed, the variance of position is
\begin{equation}
 \begin{split}
 \left<\left< X \left(t  \right)^2 \right>\right> &= 2 \gamma E \int_{0}^{t} ds_1 \int_{0}^{t} ds_2\,\delta\left( s_{1} - s_{2}  \right)  A_x\left(t-s_1 \right)  A_x\left(t-s_2 \right)  \\
 &= 2 \gamma E \int_{0}^{t} ds  A_x\left(s \right)^2,
 \end{split}
\end{equation}
where we have used the change of variable $s \to t-s$, and
\begin{equation}  \label{E_a}
\begin{split} 
E = \frac{\gamma_1 T_1 + \gamma_2 T_2}{\gamma}  + \frac{C_2}{2\gamma}.
\end{split}
\end{equation}
A similar calculation lead to
\begin{equation}
 \begin{split}
 \left<\left<  V \left( t \right)^2 \right>\right> &= 2 \gamma E \int_{0}^{t} ds A_v\left(s\right)^2, \quad \left<\left< X \left( t \right) V \left( t \right)\right>\right> 
 = 2 \gamma E \int_{0}^{t} ds\, A_x\left(s \right)A_v\left(s\right).
 \end{split}
\end{equation}
Notice that $E$ is the average energy of the harmonic oscillator in steady-state regime:
\begin{equation}
 E = \frac{m}{2}\left<\left< V^2 \right>\right> + \frac{k}{2}\left<\left< X^2 \right>\right>,
\end{equation}
\begin{equation} 
 \begin{split}
\left<\left< X^2 \right>\right> = \frac{\gamma_1 T_1 + \gamma_2 T_2}{\gamma k} + \frac{C_2}{2\gamma k}, \quad \left<\left< V^2 \right>\right> = \frac{\gamma_1 T_1 + \gamma_2 T_2}{\gamma m } + \frac{C_2}{2\gamma m}.
\end{split}
\end{equation}
For high-order cumulants of the type $\left<\left< X\left(t\right)^{n_1} V\left(t\right)^{n_2}  \right>\right>$, with $n_1,n_2 \ge 2$,  and $F$  symmetric, one can find
\begin{equation} \label{CumuGenXV1}
 \begin{split}
 \left<\left< X\left(t\right)^{n_1} V\left(t\right)^{n_2}    \right>\right> &= C_{ n_1+n_2} \left[ \prod_{i=1}^{n_1} \int_{0}^{t} ds_i A_x\left(t-s_i \right) \right] \left[ \prod_{j=1}^{n_2} \int_{0}^{t} du_j A_v\left(t-u_j \right) \right] \\
 & \quad \times \delta\left( s_1 - s_2  \right) \cdots \delta\left( s_l - u_1  \right) \cdots \delta\left( u_{n-1} - u_{n}  \right),
  \end{split}
\end{equation}
where $C_{n_1+n_2}$ is assumed to be zero for $n_1+n_2$ odd. 
The integration over time variables $s_i$ and $u_i$ are performed without difficulties; the presence of delta functions simplifies the calculation. As a result, we obtain
\begin{equation} \label{CumuXn1Vn2}
 \begin{split}
 \left<\left< X\left(t\right)^{n_1} V\left(t\right)^{n_2}  \right>\right> &= C_{n_1+n_2} \int_{0}^{t} ds A_x \left(s \right)^{n_1} A_v \left(s \right)^{n_2},
  \end{split}
\end{equation}
where $n_1$ and $n_2$ are positive integers greater than two. Observe that in the limit $\lambda \to \infty$, $b^2 \to 0$, with $\lambda b^2$ constant, we have $C_{2n} \to 0$, for $n>1$, and only the second cumulants of $X$ and $V$ are different from zero, which means that the position and velocity are Gaussian stochastic variables when $\lambda$ is very large with $\lambda b^2$ being fixed.

Since all the cumulants $\left<\left< X\left(t\right)^{n_1} V\left(t\right)^{n_2}  \right>\right>$ are determined in a closed form, we can make an attempt to find the cumulant-generating function $\mathcal{K}$ 
related to the joint distribution $P$ of $X$ and $V$:
\begin{equation}
\begin{split}
P\left( X,V,t\right) = \frac{1}{4\pi^2} \int_{-\infty}^{+\infty}\int_{-\infty}^{+\infty} d \zeta_x \, d\zeta_v \exp\left[  i\zeta_x X + i\zeta_v V +\mathcal{K}\left( \zeta_x, \zeta_v, t\right)  \right],
\end{split}
\end{equation}
where the conjugate variables $\zeta_{x,v}$ are real.
In our case---a linear stochastic system driven by independent noises---the generating function $\mathcal{K}$ is a sum of two terms, $\mathcal{K} = \mathcal{K}_1 + \mathcal{K}_2$, where
\begin{equation}  \label{CumuFuncG}
\begin{split}
\mathcal{K}_1 &= -\left( \gamma_1 T_1 +  \gamma_2 T_2 \right) \int_{0}^{t} ds \left[ 
\zeta_x A_x\left( s \right) + \zeta_v A_v\left( s \right)\right]^2,
\end{split}
\end{equation}
and
\begin{equation} \label{CumuFuncP}
\mathcal{K}_2 = \sum_{n_1,n_2} \frac{\left(-i\zeta_x \right)^{n_1}\left(-i\zeta_v \right)^{n_2}}{n_1! n_2!} C_{n_1 + n_2} \int_{0}^{t} ds A_x \left(s \right)^{n_1} A_v \left(s \right)^{n_2}.
\end{equation}
The function $\mathcal{K}_1$ presents the effects of Gaussian thermal noises, and $\mathcal{K}_2$ accounts for the contributions of non-thermal cumulants. The sums in \eqref{CumuFuncP} are allowed to run over all non-negative integer $n_i$, provided that $C_{n_1+n_2}$ is different from zero for $n_1 + n_2=2n$,  with $n \in \lbrace 1,2,3, \cdots
 \rbrace$. Due to that, the sums may be rearranged appropriately, and remembering that  $C_{2n}= \lambda b^{2n}\left( 2n-1 \right)!!$, we have
\begin{equation}
\begin{split}
\mathcal{K}_2 &= \lambda\sum_{n=1}^{\infty}  \frac{b^{2n}}{2^n n!}\left\{ \sum_{n_1=0}^{2n} \binom{2n}{n_1}  \int_{0}^{t} ds \left[-i\zeta_xA_x \left( s \right)\right]^{n_1}  \left[-i\zeta_v A_v \left( s \right)\right]^{2n-n_1} \right\}.
\end{split}
\end{equation}
The term within braces is simply a binomial expansion, which leads to
\begin{equation}
\begin{split}
\mathcal{K}_2  &= \lambda\sum_{n=1}^{\infty}  \frac{b^{2n}}{2^n n!}  \int_{0}^{t} ds \left[ -i\zeta_xA_x \left( s \right) -i\zeta_v A_v \left( s \right)\right]^{2n} \\
& = \lambda \int_{0}^t ds   \exp  \left\{   -\frac{b^2}{2} \left[   \zeta_x A_x \left( s \right) + \zeta_vA_v \left( s \right)  \right]^2 \right\} - \lambda t.
\end{split}
\end{equation}
%
%
%
Now, by using the Gaussian identity
\begin{equation}
\int_{-\infty}^{+\infty} dJ \exp\left( -\frac{J^2}{2b^2} - i\alpha J \right) = \sqrt{2\pi b^2} \exp\left( -\frac{\alpha^2 b^2}{2} \right),
\end{equation}
with $b$ and $\alpha$ being real, we can set $\alpha =  \zeta_x A_x \left( s \right) + \zeta_v A_v \left( s \right) $, which is in fact real, and obtain
\begin{equation} \label{CumuFunc1P}
\begin{split}
\mathcal{K}_2  &=  \lambda \int_{0}^t ds  \left< \exp  \left\{   -iJ \left[   \zeta_xA_x \left( s \right) + \zeta_vA_v \left( s \right)  \right] \right\} - 1 \right>_{\rho},
\end{split}
\end{equation}
where the average $\left< \cdot \right>_\rho$ is taken over the Gaussian distribution $\rho \left( J \right)$ defined in \eqref{ImpulseDist},  with zero mean and variance $b^2$. This distribution characterizes the statistics of impulse $J$ of external Poisson noise.
%
%

In addition, with the exact expressions of $\left<\left< X\left(t\right)^{n_1} V\left(t\right)^{n_2}  \right>\right>$ for given values of positive integers $n_1$ and $n_2$, the long-time limit of cumulants of position and velocity can be evaluated, as least in principle. For the case of fourth-order cumulants, a direct calculation using \eqref{CumuXn1Vn2} leads to
\begin{equation}
\left<\left< X^4 \right>\right> = \frac{3C_4}{4\gamma k \left( 3\gamma^2 + 4 km \right)}, \quad \left<\left< V^4 \right>\right> = \frac{3C_4\left(\gamma^2 + km \right)}{4\gamma m^3 \left( 3\gamma^2 + 4 km \right)},
\end{equation}
\begin{equation}
\left<\left< X^2V^2 \right>\right> = \frac{C_4}{4\gamma m \left( 3\gamma^2 + 4 km \right)}, \quad \left<\left< XV^3 \right>\right> = \frac{C_4}{4 m^2 \left( 3\gamma^2 + 4 km \right)}.
\end{equation}
The stationary cumulant $\left<\left< X^3V \right>\right>$ is zero. As a matter of fact, every cumulant of the type $\left<\left< X^{n-1}V \right>\right>$, with $n \ge 1$, tends to zero for steady-states:
\begin{equation} \label{CumuXVzero}
\begin{split}
 \left<\left< X\left(t\right)^{n-1}V\left(t\right) \right>\right>
 &= r_n \int_{0}^{t} ds A_{x}\left( s \right)^{n-1}A_{v}\left( s\right)
 = r_n  \int_{0}^{t} ds \frac{d}{ds}A_{x}\left( s \right)^{n} \\
&= r_n   A_{x}\left( t \right)^{n} \to 0 \quad \text{as} \quad t \to \infty,
\end{split}
\end{equation}
where $r_n$ is a parameter independent of time. The result \eqref{CumuXVzero} is in agreement with \cite{Morgado2015}, which used a different theoretical approach.
Indeed, it is possible to obtain an expression for the generating function \eqref{CumuFunc1P} in the limit $t \gg m/\gamma$. Consider that $\nu_{\pm} =  \nu_1 \pm i \nu_2 $ in \eqref{Para1}, which allows us to rewrite \eqref{AxAv} as
\begin{equation}
 \begin{split} 
  A_x\left( s \right) = \frac{e^{\nu_1 s}}{m \nu_2} \sin \left( \nu_2 s \right), \quad A_v\left( s \right) = \frac{e^{\nu_1 s}}{m \nu_2} \left[ \nu_1 \sin \left( \nu_2 s \right)  + \nu_2 \cos \left( \nu_2 s \right)\right].
 \end{split}
\end{equation}
Then, we set the change of variable $\mu = \exp\left( \nu_1s \right)/m\nu_2$, which is analogous to did in \cite{FodorvanWijland2018}, but with a distinct prefactor and another relaxation time-scale. Notice that $\mu$ has dimension of second over mass (i.e., reciprocal viscous coefficient). After some algebraic manipulation, we have the stationary generating function
\begin{equation} \label{StatCumuFunc1P}
\begin{split}
\mathcal{K}_{2,\text{stat}}  &=  -\frac{\lambda}{\nu_1} \int_{0}^1 \frac{d\mu}{\mu}  \left< \exp  \left\{   -iJ \mu \left[  \left( \zeta_x + \zeta_v \nu_1 \right)Z_1 \left( \mu \right) + \zeta_v \nu_2 Z_2 \left( \mu \right)  \right] \right\} - 1 \right>_{\rho},
\end{split}
\end{equation}
where
\begin{equation}
 Z_1 \left( \mu \right) = \sin \left[ \frac{\nu_2}{\nu_1} \ln \left(m\nu_2 \mu \right) \right], \quad Z_2 \left( \mu \right) = \cos \left[ \frac{\nu_2}{\nu_1} \ln \left(m\nu_2 \mu \right) \right].
\end{equation}
Notice that $\nu_{1}$ is negative. The exact expression of generating function \eqref{StatCumuFunc1P} is valid for very long times $t \gg m/\gamma$, and represents the Poisson noise effects on the steady-state statistics of position and velocity of the trapped inertial particle.

Observe that in the limit $\lambda \to \infty$, with $\lambda b^2 $ constant, we have a Gaussian distribution for $X$ and $V$ that is only consistent with the equilibrium Boltzmann-Gibbs statistics, with well-defined canonical temperature, in the absence of external noise, and with equal bath temperatures. Indeed, we will see in the next section that the stationary state of the model presents non-zero energetic currents due to external noise effects and different temperatures of thermal sources. Therefore, in our context, the external noise $F$ contributes to driving the system out of equilibrium.

\section{Stochastic heat and external work-like quantities} \label{StotchEner}

The main interest of this work is to investigate the heat exchanges and energy injected by external non-Gaussian stochastic force. For Langevin systems driven solely by Gaussian noises, it is possible to develop an 
analysis through the energetic concepts discussed in \cite{Sekimoto1998,Sekimoto2010,GomezSolano2010,PalSabhapandit2014}. In such cases, heat is identified as the work performed by the sum of friction force and thermal noise \cite{Sekimoto2010}; the energetic contribution associated with the athermal environment may be given by the work done by external noise \cite{GomezSolano2010,PalSabhapandit2014}. These work-like quantities are written as stochastic integrals of Stratonovich type \cite{Gardiner,Sekimoto2010}. In this work, for the damped linear model given by \eqref{LanSys1}, we assume that heat and external work can be interpreted analogously, despite the presence of non-Gaussian noise. We also consider calculus rules according to Stratonovich prescription.  

Within the assumption mentioned above, we introduce the dimensionless heat associated with the  thermal baths
\begin{equation} \label{StocHeat}
\begin{split} 
Q_i\left( t \right) = \frac{1}{T_{a}} \int_{0}^{t} dt_1
V\left( t_1 \right) \left[ \xi_i \left( t_1 \right) - \gamma_i V\left( t_1 \right) \right],
\end{split}
\end{equation}
and the dimensionless work-like quantity due to the non-Gaussian noise
\begin{equation} \label{StocWork}
\begin{split} 
 W\left( t \right) &= \frac{1}{T_a} \int_{0}^{t} dt_1 F\left( t_1 \right)V\left( t_1 \right).
\end{split}
\end{equation}
where 
\begin{equation}
 T_a = \frac{\gamma_1 T_1 + \gamma_2 T_2}{\gamma}.
\end{equation} 
It is also convenient to consider the additional dimensionless quantities 
\begin{equation} \label{ScaledVar1}
\widetilde{T}_i = \frac{T_i}{T_a}, \quad \eta_i = \frac{\gamma_i}{\gamma}, \quad t_o = 
\frac{ \gamma  t}{m}, \quad \Lambda=\frac{m\lambda}{ \gamma }, \quad  B = \frac{b^2}{2m T_a}.
\end{equation}
These non-dimensional variables simplify the analysis of expressions. It is straightforward to show that 
\begin{equation} \label{ScaledVar2}
\eta_1 + \eta_2 = 1, \quad \eta_1 \widetilde{T}_1 + \eta_2\widetilde{T}_2 = 1.
\end{equation}
The case with equal bath temperatures $T_i = T$ leads to $\widetilde{T_i} =1$ and $B =  b^2/2mT$. For external Gaussian noise, which can be obtained by taking $\Lambda \to \infty $, $B \to 0$, with $\Lambda B =  \Gamma = const.$, the quantity $\Gamma $ is the only relevant parameter associated with athermal medium. By analogy with the second cumulants of thermal noises, we may interpret $\Gamma$ as proportional to an dimensionless temperature-like quantity related to athermal reservoir. When the external noise is of Poisson kind, its stochastic properties depend on the parameters $\Lambda$ and $B$.

According to our formulation, the energetic exchanges $Q_i$ and $W$ are associated with different surroundings. Due to that, we understand it is reasonable to consider different quantities for each energetic contribution. Notice that $W$ and $Q_i$ are correlated because the system---a Brownian particle trapped in a harmonic potential---is simultaneously coupled to different reservoirs.

Indeed, we may obtain information about the fluctuating behavior of $Q_i$ and $W$ by evaluating their cumulants, as long as such calculation is possible. Then, by assuming the moments are feasible to determine, we follow a treatment analogous to the one discussed in
\cite{MorgadoQueiros2014,MorgadoQueiros2016} in order to calculate the cumulants of energetic transfers. According to \eqref{StocHeat} and \eqref{StocWork}, $Q_i$ and $W$ are given in terms of the instantaneous velocity $V\left(t \right)$ and noises $\xi_i\left( t \right)$ and $F\left( t \right)$. However, the velocity \eqref{XeqVeq} is a superposition of independent stochastic forces. This allows us to write the moments of $Q_i$ and $W$ through time integrals involving moments of uncorrelated noises. We expand the moments of noises into cumulants, and the integrals are evaluated for very long times. With the expressions for the moments, we determine the long-term cumulants of $Q_i$ and $W$. 

\subsection{Fluctuations of energy exchanged with athermal source: Gaussian case} \label{FlucWGauss}

For the limit of very small intensities $b$ and large Poisson rate $\lambda$, with $\lambda b^2 = C_2$ fixed, i.e, $ \Lambda B = \Gamma = const.$, the athermal noise $F$ approaches a Gaussian stochastic force. This case is investigated in \cite{Sabhapandit2011} with a single thermal noise. Nonetheless, in this work, we are supposing the action of two distinct heat sources and an external Gaussian process. The cumulants of $W$ can be determined by calculating its moments. The general expression of moment of order $n$ of $W$ is given by
\begin{equation} \label{AvgWn}
\begin{split}
 T_a^n \left< W\left( t \right)^n \right> &= \left<  \prod_{j=1}^n\int_{0}^{t}dt_{j} \int_{0}^{t_j}ds_{j}\, A_{v}\left( t_j - s_j \right) F\left( t_j \right)\left[ F\left(s_j \right) + \xi \left( s_j \right) \right] \right>.
\end{split}
\end{equation}
This equation is valid for $F$ being a Gaussian or Poissonian stochastic force. Then, we write the moments in terms of cumulants;
the integrals are performed and the main contributions for $t$ very long are considered. As a result, for $F$ Gaussian, one can find
\begin{equation}
T_a\left< W\right> = \frac{C_2 t}{2m},
\end{equation}
\begin{equation}
 T_a^2\left< W^2 \right> =  
 \frac{C_2^2 t^2}{4 m^2} + \left( t\gamma -m \right) \frac{C_2 E}{m\gamma},
\end{equation}
\begin{equation}
\begin{split}
T_a^3 \left< W^3 \right> &= \frac{C_2^3 t^3}{8 m^3} + \left( t^2\gamma^2 + t\gamma m -4m^2 \right)\frac{3C_2^2 E}{2m^2\gamma^2},
\end{split}
\end{equation}
\begin{equation}
\begin{split}
T_a^4 \left< W^4 \right> &= \frac{C_2^4 t^4}{16 m^4} + \frac{3C_2^3t^3 E }{2m^3} +  \left( 3C_2 + 2\gamma E \right)\frac{3C_2^2t^2 E}{2m^2 \gamma} - \left( 6C_2 + 2\gamma E \right)\frac{6C_2^2 E}{\gamma^3}.
\end{split}
\end{equation}
where $E$ is the average energy of the system shown in \eqref{E_a}. Now, using the relations between moments and cumulants \cite{van_Kampen-book,Gardiner} and the dimensionless quantities \eqref{ScaledVar1}, it is possible to show that
\begin{equation} \label{CumuJGauss1}
 \begin{split}
  \frac{1}{t_o} \left<\left<  W\right>\right>= \Gamma, \quad \frac{1}{t_o} \left<\left<  W^2 \right>\right>= 2\Gamma \left( 1 + \Gamma \right),
 \end{split} 
\end{equation}
\begin{equation} \label{CumuJGauss2}
 \begin{split}
  \frac{1}{t_o} \left<\left<  W^3 \right>\right>= 12\Gamma^2 \left( 1 + \Gamma \right), \quad \frac{1}{t_o} \left<\left<  W^4 \right>\right>= 24 \Gamma^2\left( 1 + \Gamma \right)\left( 1 + 5\Gamma \right).
 \end{split} 
\end{equation}
Although the system is coupled to two thermal baths and one athermal reservoir, the cumulants \eqref{CumuJGauss1}--\eqref{CumuJGauss2} are written only in terms of the dimensionless quantity $\Gamma$. Then, if we consider the particular case where there exists just a single heat reservoir, for example, by taking $\eta_2, \widetilde{T}_2 \to 0$, the first four cumulants of $W$ are still of the form \eqref{CumuJGauss1}--\eqref{CumuJGauss2} by adopting the dimensionless parameters \eqref{ScaledVar1}.

Observe that, according to \eqref{CumuJGauss1}--\eqref{CumuJGauss2}, the long-time limit of the distribution function of $W$ is not Gaussian: high-order cumulants beyond the second are different from zero. Indeed, the athermal energy $W$ is proportional to the time integral of the power $\mathcal{P}_{ex} \left(t \right) = F\left( t\right) V\left( t\right)$, which is a product of Gaussian variables when external noise is Gaussian, and that type of product usually is not Gaussian. Also, the autocorrelation function of $\mathcal{P}_{ex}$ is 
\begin{equation} \label{CorrPexGauss}
 \begin{split}
 \left<\left< \mathcal{P}_{ex} \left(t_1 \right) \mathcal{P}_{ex} \left(t_2 \right)\right> \right> &= \left<\left< F\left( t_1\right) F\left( t_2 \right) \right> \right>  \left<\left< V\left( t_1\right)V\left( t_2 \right) \right> \right> \\
 & \quad + \left<\left< F \left( t_1\right) V\left( t_2\right) \right> \right>\left<\left< F \left( t_2\right) V\left( t_1\right) \right> \right>,
  \end{split}
\end{equation}
with $\left<\left< F \left( t_1\right) V\left( t_2\right) \right>\right> = C_2 H\left( t_2 - t_1 \right) A_{v}\left( t_2 - t_1\right)$, 
where $H\left( t \right)$ is the Heaviside step function and $A_{v}\left(t \right)$ is shown in \eqref{AxAv}. The velocity-velocity autocorrelation function in steady-state regime is given by $\left<\left< V\left( t_1 \right) V\left( t_2 \right) \right>\right> = E A_{v} \left( \lvert t_1 - t_2\rvert \right)$, which decays exponentially in time difference $t_1 - t_2$.
As a result, it is possible to perceive that \eqref{CorrPexGauss} is zero for $t_1 \neq t_2$, but it is diverges when $t_1 = t_2$ because $\left<\left< F\left( t\right)^2 \right> \right>$ is not well-defined in the white noise limit (see \eqref{NoiseCumuPoisson}). In fact, the variance of $\mathcal{P}_{ex}$ is singular when the noise $F$ is white, which is also obtained in \cite{MorgadoQueiros2014,MorgadoQueiros2016} for the case of a harmonic oscillator driven by a single white noise. A sum of stochastic variables, each one with infinity variance, does not fulfill the conditions of the Central Limit Theorem \cite{Gardiner}. Thus, it is reasonable that $W$ does not follow Gaussian statistics.

As we said before, when there is a single thermal reservoir (e.g., as $\eta_2,\widetilde{T}_2 \to 0$), in addition to an athermal Gaussian noise, the expressions of cumulants are still the same as shown in \eqref{CumuJGauss1}--\eqref{CumuJGauss2}, but with $\eta_1 = 1$ and $\widetilde{T_i}=1$, since we are working with non-dimensional quantities. In this case, our calculation of the first cumulants of $W$ are in agreement with results discussed in \cite{Sabhapandit2011,Sabhapandit2012} for a inertial Brownian particle trapped in an harmonic well and subjected to thermal and athermal  Gaussian stochastic forces. On the other hand, when the effects of external $F$ are dominant (i.e., for $\Gamma \gg 1$), the cumulant-generating function of $W$, which is of the form $\mathcal{Y} \left( z \right) = \ln \left< \exp \left( z W \right) \right>$, can be written as a series expansion
\begin{equation}
 \begin{split}
  \frac{1}{t_o} \mathcal{Y} \left( z \right) &= 
  \frac{\Gamma z}{1!} +  \frac{2 \Gamma ^2 z^2}{2!} + \frac{12 \Gamma^3 z^3}{3!} + \frac{120 \Gamma^4 z^4}{4!} + \cdots .
 \end{split}
\end{equation}
Then, by assuming this series converges, we may try to rearrange the first terms in order to see if some elementary functions can be identified. In fact, this is possible and gives
\begin{equation} \label{CumuGenFuncWGauss}
 \begin{split}
  \frac{1}{t_o} \mathcal{Y} \left( z \right)  &= \frac{1}{2} \left[ \frac{4\Gamma z}{2^1 \cdot 1!} +  \frac{\left( 4\Gamma z\right)^2}{2^2 \cdot 2!} + \frac{3\left( 4\Gamma z\right)^3}{2^3 \cdot 3!} + \frac{15\left( 4\Gamma z\right)^4}{2^4 \cdot 4!} + \cdots \right] \\
 &= \frac{1}{2} \left\{  1 -\left[ 1 + \frac{\left( -4\Gamma z \right)}{2^1 \cdot 1!} -  \frac{\left( -4\Gamma z\right)^2}{2^2 \cdot 2!} + \frac{3\left( -4\Gamma z\right)^3}{2^3 \cdot 3!} - \frac{15\left( -4\Gamma z\right)^4}{2^4 \cdot 4!} + \cdots \right] \right\} \\
  &= \frac{1}{2} - \frac{1}{2}\sqrt{1 - 4\Gamma z}.
 \end{split}
\end{equation}
which is consistent with the analysis presented in \cite{Farago2002}.

\subsection{First cumulants of energy injected by Poisson noise} \label{FlucWPoisson}

When the Poisson rate $\lambda$ and variance of impulse distribution $b^2$ are finite, the system is under the influence of a non-Gaussian stochastic force. In this case, by considering that $t$ is very long, time-independent contributions can be disregarded in calculating the moments of $W$. Consequently, we find
\begin{equation}
 T_a\left< W\right> = \frac{C_2 t}{2m},
\end{equation}
\begin{equation}
 T_a^2 \left< W^2 \right> = \frac{C_4 t}{4m^2} + \frac{C_2^2 t^2 }{4m^2} + \left(   C_2 + 2\gamma_1 T_1 + 2\gamma_2 T_2 \right)\frac{ \left( t \gamma  -m \right) C_2}{ 2m \gamma^2},
\end{equation}
\begin{equation}
 \begin{split}
 T_a^3 \left< W^3 \right> &= \frac{C_6 t}{8m^3} + \frac { 3\left( t \gamma  -m \right) }{  2 
 \left(m \gamma \right)^2 } \left( {\gamma_1}{T_1}+{\gamma_2}{T_2}
 \right)   C_4 + \frac{ 3\left[ \left( t \gamma \right)^{2} +4t \gamma m -4{m
}^{2} \right] }{ 8{m}^{3}{{\gamma}}^{2}} {C_4}{C_2}\\
& \quad + \frac{ 3\left[ \left(t \gamma \right)^2 +t \gamma m -4{m
}^{2} \right] }{ 2m^2\gamma^3 }  \left( {\gamma_1}{T_1}+{\gamma_2}{T_2}
 \right) C_2^2 \\
 & \quad +  \frac{  \left[ \left(t \gamma \right)^3 -24{m}^{
3}+6\left(t \gamma \right)^2 m+6\gamma t{m}^{2} \right]  }{ 8\left(m \gamma \right)^3 }C_2^3.
  \end{split}
\end{equation}
The cumulants are obtained through the expressions $\left<\left< W \right>\right> = \left< W \right>$, $\left<\left< W^2 \right>\right> = \left<W^2 \right> - \left< W\right>^2$ and $\left< \left< W^3 \right>\right> = \left<W^3 \right> - 3\left< W\right>\left< W^2 \right> + 2 \left< W\right>^3$, which lead to
\begin{equation} \label{CumuJ}
\begin{split}
 \frac{ 1 }{t_o} \left< \left< W \right> \right>&= \Lambda B ,
\end{split}
\end{equation}
\begin{equation} \label{CumuJ2}
\begin{split}
 \frac{ 1 }{t_o} \left< \left< W^2 \right> \right>&= \Lambda B \left[ 2\left( 1 + \Lambda B \right) + 3B \right],
\end{split}
\end{equation}
\begin{equation} \label{CumuJ3}
\begin{split}
 \frac{ 1 }{t_o} \left< \left< W^3 \right> \right>= 6\Lambda B^2 \left( 3 + 2\Lambda\right) + 3\Lambda B^3 \left( 5 + 2\Lambda\right)\left( 1 + 2\Lambda\right).
\end{split}
\end{equation}
where the dimensionless quantities \eqref{ScaledVar1} are adopted.
Observe that, for $F$ being Poissonian \cite{van_Kampen-book}, the quantity $\lambda t = \Lambda t_o$ is the average number of impulses in the interval $t$. As a result, we may interpret $B$ as the mean dimensionless energy per impulse injected by the Poisson noise.

We also evaluate the averages of products of $W$ and $Q_i$:
\begin{equation} \label{AvgQiW}
 \begin{split}
  T_a^2\left< Q_i \left( t \right) W\left( t \right)  \right> &= \int_{0}^t dt_1 \int_{0}^t dt_2 
  \left<  \xi_i \left( t_1 \right) V\left( t_1 \right) F\left( t_2 \right)V\left( t_2 \right) \right> \\
  & \quad -\gamma_i \int_{0}^t dt_1 \int_{0}^t dt_2 
  \left<  F\left( t_2 \right) V\left( t_2 \right) V\left( t_1 \right)^2 \right>.
 \end{split}
\end{equation}
For very long times ($t \gg m/\gamma$), we have 
\begin{equation}
 \begin{split}
 T_a^2 \left<  Q_i W\right> &= - \frac{\left( t \gamma -m\right) \gamma_i C_4}{ \left( m \gamma \right)^2} + \frac{ t^2  \gamma_i T_i C_2}{2m^2} - \frac{ \left[ \left( t \gamma \right)^{2} + t \gamma m -4{m
}^{2} \right] }{ 2m^2\gamma^2 }\gamma_i E C_2.
 \end{split}
\end{equation}
Then, in terms of non-dimensional quantities, the covariances $\left<\left< Q_i W\right>\right> = \left<Q_i W\right> - \left< Q_i \right>\left< W\right>$ can be written as 
\begin{equation} \label{CumuQiJ}
 \frac{ 1 }{t_o}\left<\left<   Q_i W \right>\right>   = -\eta_i \Lambda B\left[ 2\left( 1 + \Lambda B \right) + 3B \right].
\end{equation}

We see that the average injected energy (fisrt cumulant $\left< \left< W \right> \right>$), the standard deviation (square root of $\left< \left< W^2 \right> \right>$) and the asymmetry 
(third cumulant $\left< \left< W^3 \right> \right>$) of the distribution of $W$ depend on the dimensionless quantities $\Lambda$ and $B$. These parameters are proportional to the Poisson rate $\lambda$ and impulse variance $b^2$, which are related to external noise cumulants \eqref{NoiseCumuPoisson}--\eqref{EqC2n}. Then, the quantities $\left<\left< W^n \right>\right>$ are influenced by the cumulants of $F$, i.e., the non-Gaussian fluctuations due to external noise contribute to the fluctuations of energy injection $W$. This may be important for the energetic considerations of mesoscopic systems driven by external stochastic forces, since the nature of noise affects the statistics of energy exchanges with the reservoirs. For example, in the context of microsized engines investigated experimentally in \cite{DinisRica2016,RoyGanapathy2021}.
Notice that the covariances $\left<\left< Q_i  W\right>\right>$ are always less than or equal to zero, as well as proportional to $\left<\left<  W^2 \right>\right>$, where the coefficient of proportionality is $\eta_i$, which is related to the friction coefficient that represents the coupling to thermal bath. It is straightforward to check that in the limit of very high rate $\Lambda$ and $\Lambda B = \Gamma $ fixed, one can recover the cumulants \eqref{CumuJGauss1}--\eqref{CumuJGauss2} for the case of external Gaussian noise. On the other hand, in the limit of small temperatures ($B$ large), and considering $\eta_i  \neq 0$, the expressions for $\left<\left< W^n \right>\right>$, $n \in \left\{ 1,2,3\right\}$, written in appropriate units, are in agreement with \cite{MorgadoQueiros2016}.

\begin{figure}
 \centering
 \includegraphics[scale=0.5]{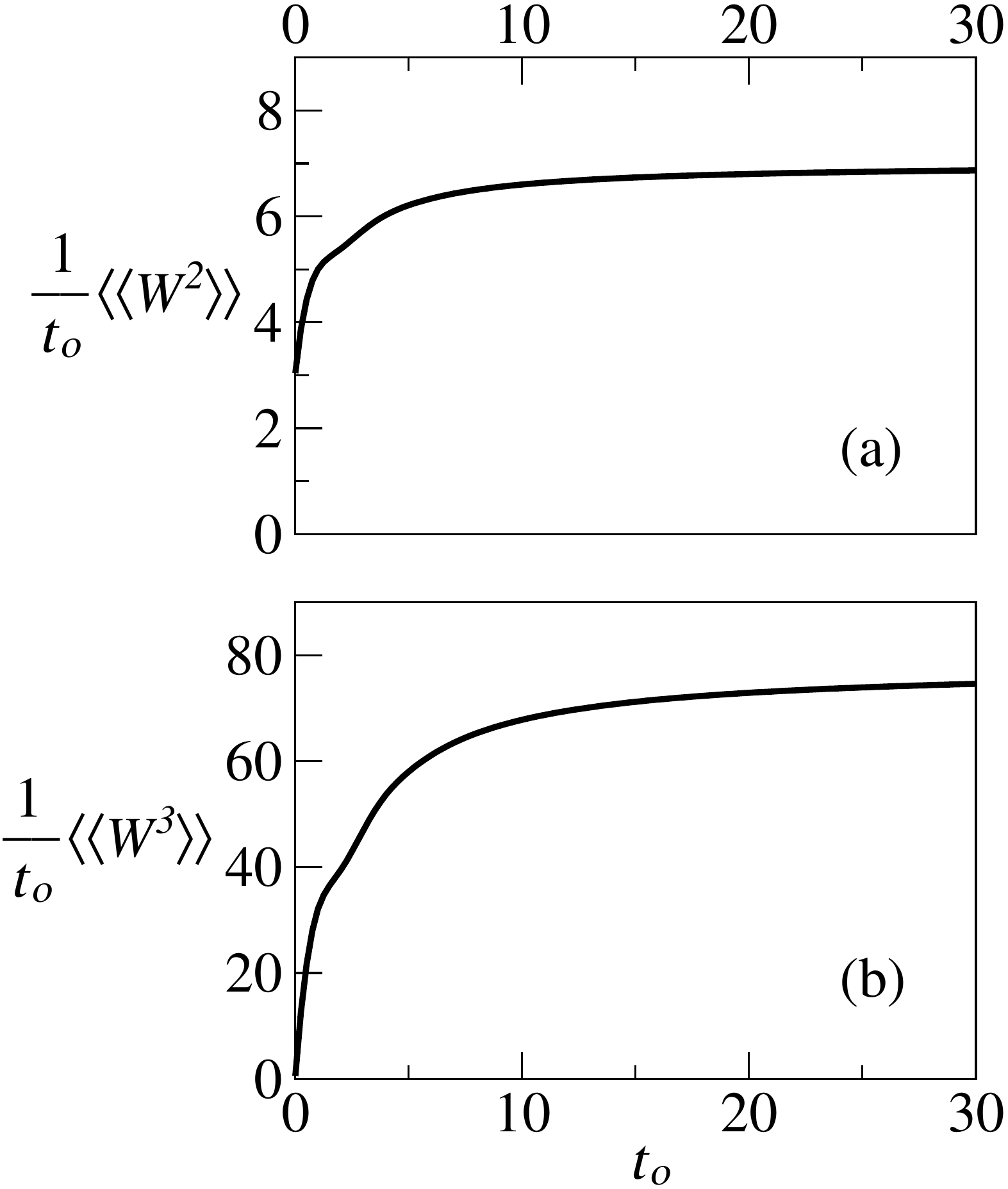}
 \caption{$\left<\left< W^2 \right>\right>/t_o $ and $ \left<\left< W^3 \right>\right>/t_o$ versus the dimensionless time $t_o$ for $\Lambda=1$, $B=1$, and $\Upsilon = 2$. When $t_o \gg 1$, the cumulants achieve a long-time behavior where $\left< \left< W^n  \right> \right>$ are proportional to $t_o$.} 
\label{W2W3time}
\end{figure}

Although we focus on the long-time limit, the expression of second cumulant of $W$ is feasible to be shown for any value of $t_o$:
\begin{equation} \label{W2_finite_t0}
 \begin{split}
 \left< \left< W^2 \left( t_o \right) \right> \right> &= \Lambda B\left( 2 + 3B + 2\Lambda B \right)t_o \\
 & \quad + \frac{2 \Lambda B}{\Upsilon^2} \left[ e^{-t_o}\left( \Upsilon^2 + 1 - \cos \Upsilon t_o \right) - \Upsilon^2 \right],
 \end{split}
\end{equation}
where 
\begin{equation} \label{Ypara}
 \Upsilon = \sqrt{ \frac{4km}{\gamma^2} -1 }, \quad 4km>\gamma^2.
\end{equation}
The first term in \eqref{W2_finite_t0} is proportional to $t_o$, which leads to the main contribution when $t_o \gg 1$. There is also a term that decays exponentially with time and only contributes in a significant way in the transient regime. Notice that \eqref{W2_finite_t0} does not depend explicitly on $\eta_i$ and $T_i$. The equation for the instantaneous third cumulant $\left< \left< W^3 \left( t_o \right) \right> \right>$ is long and complicated to be written here, but it also depends only on $\Lambda$ and $B$, with one term linear in $t_o$ and another that vanishes exponentially for $t_o \gg 1$. We show in Fig. \ref{W2W3time} the time dependence of second and third cumulants of $W$, where it is possible to identify the evolution to a long-term regime in which $\left< \left< W^n \right> \right>$ scale with $t_o$. It is worth mentioning that the first cumulant of $W$ is given by \eqref{CumuJ} for any $t_o$ positive.

The presence of non-zero third cumulant $\left<\left< W^3 \right>\right>$ (i.e. the skewness) is an indication of non-Gaussian behavior related to $W$. In fact, the stochastic power $\mathcal{P}_{ex} = F \left( t\right) V\left( t\right)$ is a product of Gaussian and non-Gaussian variables, and probably is non-Gaussian. In addition, for an external Poissonian $F$, the autocorrelation function of $\mathcal{P}_{ex}$ is given by 
\begin{equation} \label{CorrPexPoisson}
 \begin{split}
 \left<\left< \mathcal{P}_{ex} \left(t_1 \right) \mathcal{P}_{ex} \left(t_2 \right)\right> \right> &= \frac{C_4}{2m} \delta\left( t_1 - t_2 \right)H\left( t_1 - t_2 \right)A_v\left( t_1 - t_2 \right) \\
 & \quad + \left<\left< F\left( t_1\right) F\left( t_2 \right) \right> \right>  \left<\left< V\left( t_1\right)V\left( t_2 \right) \right> \right> \\
 & \quad + \left<\left< F \left( t_1\right) V\left( t_2\right) \right> \right>\left<\left< F \left( t_2\right) V\left( t_1\right) \right> \right>,
  \end{split}
\end{equation}
where $H\left( t \right)$ is the Heaviside step function. 
The first term in \eqref{CorrPexPoisson} is due to the non-Gaussian nature of athermal noise and tends to zero in the Gaussian limit of $F$, which leads to the particular case \eqref{CorrPexGauss}. Clearly, the autocorrelation function $\left<\left< \mathcal{P}_{ex} \left(t_1 \right) \mathcal{P}_{ex} \left(t_2 \right)\right> \right>$ as well as the variance $\left<\left< \mathcal{P}_{ex} \left(t\right)^2 \right> \right>$ are singular quantities as long as $F$ is white. This is in agreement with the results found in \cite{MorgadoQueiros2016}
for a single stochastic force of Poisson kind. Therefore, the time-integrated variable $W$ is not expected to be Gaussian. Observe that, according to results discussed in \ref{FlucWGauss}, the fluctuations of $W$ are also non-Gaussian when $F$ is a Gaussian noise. Nevertheless, it is important to emphasize that, for $F$ being a Poisson-type noise, its non-Gaussian aspects influence $W$ in a distinct way, as indicated by cumulants shown in \eqref{CumuJ}-\eqref{CumuJ3}.

\subsection{Averages and covariances of heat}

\begin{figure}
 \centering
 \includegraphics[scale=0.5]{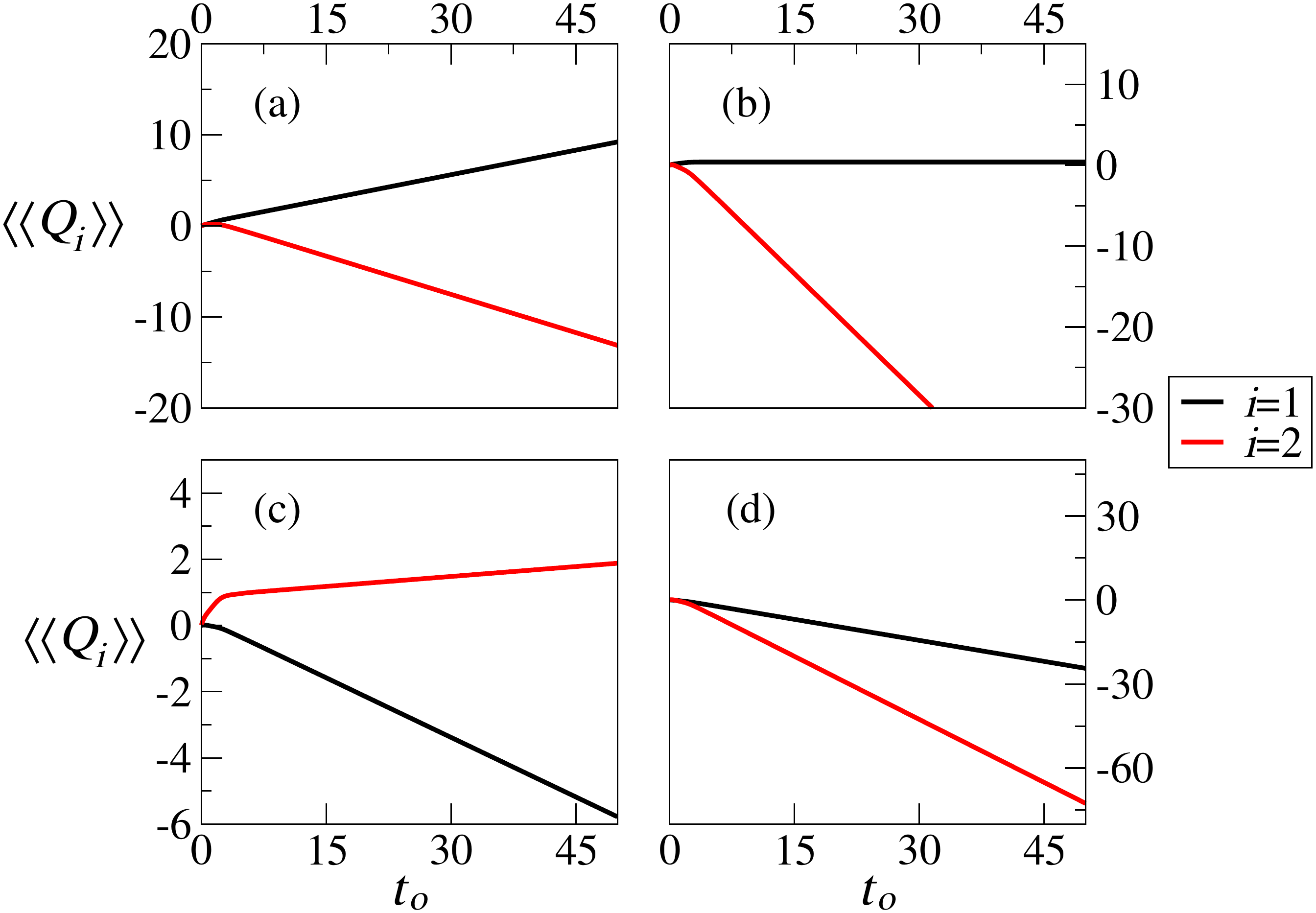}
 \caption{ Average heat $\left<\left< Q_i \right>\right>$ versus dimensionless time $t_o$ with $\eta_1=0.2$, $\Upsilon = 2$, (a) $\widetilde{T}_1 = 2$, $\Lambda=0.1$, $B = 1 $; (b) $\widetilde{T}_1 = 2$, $\Lambda=1$, $B = 1 $; (c) $\widetilde{T}_1 = 0.5$, $\Lambda=0.1$, $B = 1 $; and (d) $\widetilde{T}_1 = 0.5$, $\Lambda=2$, $B = 1 $. The values of $\eta_2$ and $\widetilde{T}_2$ are given by \eqref{ScaledVar2}.}
\label{AvgQi}
\end{figure}

Again, using the formal solutions of Langevin-like system \eqref{XeqVeq} and the cumulants of noises \eqref{NoiseCumuGauss}--\eqref{NoiseCumuPoisson}, we obtain the averages heat
\begin{equation} \label{InstQ1}
 \begin{split}
 \left<  Q_1 \left( t \right) \right> &= \eta_1 \eta_2\left( \widetilde{T}_1 - \widetilde{T}_2 \right)t_o - \eta_1\Lambda Bt_o + 1 + \Lambda B \\
 & \quad + \frac{\eta_1 \left( 1 + \Lambda B\right)}{\Upsilon^2} e^{-t_o} \left\{  2\left[ \cos\left( \frac{t_o \Upsilon}{2} \right)\right]^2 - 2 - \Upsilon^2 \right\},
 \end{split}
\end{equation}
\begin{equation} \label{InstQ2}
 \begin{split}
 \left<  Q_2 \left( t \right) \right> &= \eta_1 \eta_2\left( \widetilde{T}_2 - \widetilde{T}_1 \right)t_o - \eta_2\Lambda Bt_o + 1 + \Lambda B \\
 & \quad + \frac{\eta_2 \left( 1 + \Lambda B\right)}{\Upsilon^2} e^{-t_o} \left\{  2\left[ \cos\left( \frac{t_o \Upsilon}{2} \right)\right]^2 - 2 - \Upsilon^2 \right\}
 \end{split}
\end{equation}
where $\Upsilon$ is defined in \eqref{Ypara}. The time-dependent averages $\left< Q_i \left(t \right) \right>$ present terms linear in $t_o$, which is the relevant contribution for very long times, terms that becomes exponentially small for $t_o \gg 1$, in addition to terms that do not depend on time. We show in Fig. \ref{AvgQi} graphs of the averages of $Q_i$ against $t_o$ for different values of dimensionless  parameters. In the long-time limit ($t \gg m/\gamma$), the first cumulants of $Q_i$ are given by
\begin{equation} \label{Cumu1Qa}
 \frac{ 1 }{t_o} \left< \left< Q_1 \right> \right>  = \eta_1 \eta_2 \left( \widetilde{T}_1 - \widetilde{T}_2 \right)  - \eta_1 \Lambda B,
\end{equation}
\begin{equation} \label{Cumu1Qb}
 \frac{ 1 }{t_o} \left< \left< Q_2 \right> \right>  = \eta_1 \eta_2 \left( \widetilde{T}_2 - \widetilde{T}_1 \right)  - \eta_2 \Lambda B.
\end{equation}
Roughly speaking, the averages $\left<\left< Q_i \right>\right>$ present some similarities with the macroscopic thermal exchanges of energy between the system and heat baths. Indeed, let us consider the steady-state mean currents $h_i$ and $w$ such as $\left<\left< Q_i \right>\right> = h_i t_o + \text{const}$ and $\left<\left< W\right>\right> = w t_o + \text{const}$, with $t\gg m/\gamma$.
These average energetic currents lead to a kind of first law form (conservation of energy): $\left<\left< W\right>\right> + \left<\left< Q_1 \right>\right> + \left<\left< Q_2 \right>\right> = E/T_a$, which gives $h_1 + h_2 + w=0$.
Now, suppose that  $\eta_2 \left( \widetilde{T}_1 - \widetilde{T}_2\right) > \Lambda B$, with $\widetilde{T}_1>\widetilde{T}_2$. Then, we have $h_1>0$ and $h_2<0$, which means that heat is flowing through the system from reservoir $1$ to bath $2$. Otherwise, if $\eta_2 \left( \widetilde{T}_1 - \widetilde{T}_2\right) < \Lambda B$, but keeping $\widetilde{T}_1>\widetilde{T}_2$, we find $h_1 <0$ and $h_2 <0$, which show that heat is delivered to thermal baths. As a result, we have two scenarios: there exists average heat flowing  through the system from hotter 
bath to colder one, or thermal environments absorb (on average) the energy injected by 
external noise.

For the particular case with a single thermal bath, for example, if $\eta_2=0$ and $\widetilde{T}_2 = 0$, there is a current of energy from athermal source to heat reservoir: $h_1 =-\Lambda B$, $w = \Lambda B$.
Here, it is possible to perceive an interesting connection with the use of electric shaking as a heating mechanism in~\cite{Rossnagel2016}. In this case, energy in the form of heat, which originally comes from the external work done upon the system, is pumped into the thermal reservoir---exactly as another source of heat at higher temperature would have done. Indeed, for mesoscopic fluctuating engines, external noises can be used to design reservoir-like effects \cite{DinisRica2016,RoyGanapathy2021}.
Then, the role of Gaussian and non-Gaussian external driving can be a point of interest. 
Observe that, for $F$ Poissonian and also considering $\eta_2$ and $T_2$ zero, the average heat current $-\Lambda B$ dissipated in the thermal bath depends on the rate of kicks $\Lambda$ and the mean dimensionless injected energy $B$, both accessible experimentally in principle. In addition, according to \eqref{CumuQiJ}, $Q_1$ and $W$ are negatively correlated, where the covariance $\left<\left< Q_1 W\right>\right>$ decreases as $\Lambda$ and $B$ increase. Alternatively, by assuming the change of variable $\Lambda B = \Gamma$, we have $\left<\left< Q_1 \right> \right>/T_o = -\left<\left< W \right> \right>/T_o=-\Gamma$ and $\left< \left< W^2 \right> \right>/t_o = -\left< \left< Q_1 W \right> \right>/t_o = \Gamma \left[ 2\left( 1 + \Gamma \right) + 3\Gamma/\Lambda \right]$. As a result, the average currents $\left< \left< Q_1 \right> \right>/t_o$ and $\left< \left< W \right> \right>/t_o$ present the same form for Gaussian ($\Lambda \gg 1$, $\Gamma$ finite) and Poissonian noises, but the expressions of $\left< \left< W^2 \right> \right>$ and $\left< \left< Q_1W \right> \right>$ depend on the noise considered. This show the influence of the nature of the external stochastic force on the fluctuations of energy transfers.

It is also important to consider the effects of external noise on the covariances of 
$Q_i$.
The general expression of second moments of heat is
\begin{equation} \label{qiqj}
 \begin{split}
 T_a^2 \left< Q_i \left( t \right) Q_j \left( t \right) \right> &=  \int_0^t dt_1 \int_0^t dt_2 \Big\{  \left<  \xi_i\left( t_1 \right)  V \left( t_1  \right) \xi_j\left( t_2 \right)  V\left( t_2  \right) \right>   \Big. \\
 & \quad + \gamma_i \gamma_j  \left< V \left( t_1 \right)^2 V \left( t_2 \right)^2 \right>  - \gamma_i \left< \xi_i\left( t_1 \right)  V \left( t_1  \right)  V \left( t_2 \right)^2 \right> \\
 & \quad \Big. - \gamma_j\left< \xi_j\left( t_2 \right)  V \left( t_2  \right)
  V \left( t_1 \right)^2 \right>  \Big\}.
  \end{split}
\end{equation}
We determine each contribution of \eqref{qiqj} in a similar way to did in \ref{FlucWGauss} and \ref{FlucWPoisson}. Then, after performing the cumulant expansion of moments of noises as well as the time integrals, we find
\begin{equation} 
 \begin{split}
 \left< Q_i Q_j \right> = \frac{I_{ij} +\gamma_i \gamma_j \left(
  I^{\prime} -  I_{ij}^{\prime} \right) }{T_a^2},
 \end{split}
\end{equation}
where
\begin{equation} 
 \begin{split}
 I_{ii} = \left( \frac{ t  \gamma_i T_i }{m}\right)^2 +   \frac{2\left( t \gamma  - m \right)\gamma_i T_i E}{m \gamma}, \quad I_{ii}^{\prime} =\frac{ 2\left[ \left( t \gamma  + m \right)t \gamma - 4m^2 \right]   T_i E}{ \left( m\gamma \right)^2  } ,
 \end{split}
\end{equation}
\begin{equation} 
 \begin{split}
  I^{\prime} &= \frac{  C_4 \beta }{16m^2 \gamma^3  \left( 3{{\gamma}}^{2}+4km \right)  }   +  \frac{ \left[ \left( t \gamma \right)^2 -4m^2   \right]E^2}{ \left( m \gamma \right)^2 },
 \end{split}
\end{equation}
\begin{equation}
\beta = 4\left( 3\gamma^2 + 4 km \right)\gamma t - 3m \left( 7\gamma^2 + 8 km \right),
\end{equation}
\begin{equation} 
 \begin{split}
 I_{12} = I_{21} =  \frac{t^2 \gamma_1 \gamma_2  }{m^2} T_1 T_2, 
  \quad I_{12}^{\prime} = I_{21}^{\prime} = \frac{ I_{11}^{\prime} + I_{22}^{\prime} }{2}.
 \end{split}
\end{equation}
with $E$ given by \eqref{E_a}. These expressions allow us to write the cumulants $ \left<\left< Q_i Q_j \right>\right> = \left< Q_i Q_j \right> - \left< Q_i \right> \left< Q_j\right>$.
%
%
%
%
As a result, we obtain the variances of $Q_1$ and $Q_2$:
\begin{equation} \label{Cumu2Q1}
\begin{split}
 \frac{ 1 }{t_o} \left< \left< Q_1^2 \right> \right>  &=  2\eta_1 \eta_2 \left( \eta_2 \widetilde{T}_1 + \eta_1 \widetilde{T}_2 \right) + 2\eta_1 \Lambda B \left[ \left( \eta_1^2 + \eta_2^2 \right) \widetilde{T}_1 + 2 \eta_1 \eta_2 \widetilde{T}_2 \right] \\
 & \quad   + \eta_1^2 \Lambda B^2 \left( 3 + 2\Lambda \right),
\end{split}
\end{equation}
\begin{equation} \label{Cumu2Q2}
\begin{split}
 \frac{ 1 }{t_o} \left< \left< Q_2^2 \right> \right>  &=  2\eta_1 \eta_2 \left( \eta_2 \widetilde{T}_1 + \eta_1 \widetilde{T}_2 \right) + 2\eta_2 \Lambda B \left[ \left( \eta_1^2 + \eta_2^2 \right) \widetilde{T}_2 + 2 \eta_1 \eta_2 \widetilde{T}_1 \right] \\
 & \quad   + \eta_2^2 \Lambda B^2 \left( 3 + 2\Lambda \right).
\end{split}
\end{equation}
The covariance of $Q_1$ and $Q_2$ are given by 
\begin{equation} \label{CumuQ1Q2}
\begin{split}
 \frac{ 1 }{t_o} \left< \left< Q_1Q_2 \right> \right>  &=  -2\eta_1 \eta_2 \left( \eta_2 \widetilde{T}_1 + \eta_1 \widetilde{T}_2 \right) + 2\eta_1 \eta_2 \Lambda B \left( \eta_1 - \eta_2 \right) \left( \widetilde{T}_1- \widetilde{T}_2 \right)  \\
 & \quad   + \eta_1 \eta_2 \Lambda B^2 \left( 3 + 2\Lambda \right).
\end{split}
\end{equation}
Observe that the second cumulants of $W$ and $Q_i$ are elements of the covariance matrix associated with the joint distribution of $W$ and $Q_i$. 

The variances of $Q_1$ and $Q_2$ are influenced by the properties of external noise, as well as the temperatures of thermal reservoirs and friction coefficients. The same is true for the second cumulant $\left<\left< Q_1 Q_2 \right>\right>$. Notice that the quantity $\left<\left< Q_i^2 \right>\right>/2t_o$ can be interpreted as the diffusion coefficient associated with the stochastic flow of heat $Q_i$. Analogously, $\left<\left< W^2 \right>\right>/2t_o$ would be the diffusion coefficient related to $W$. By neglecting the effects of external noise (i.e. $\Lambda, B \to 0$), one can obtain
\begin{equation}
\frac{1}{t_o}\left< \left< Q_i^2 \right>\right> \to 2\eta_1 \eta_2 \left( \eta_2\widetilde{T}_1 +  \eta_1\widetilde{T}_2 \right),
\end{equation}
which is in agreement with \cite{FogedbyImparato2011} by using the non-dimensional variables \eqref{ScaledVar1}. In the Gaussian limit of external noise ($\Lambda \to \infty$, $\Lambda B = \Gamma $ constant), we have
\begin{equation}
\frac{ 1 }{t_o} \left< \left< Q_1^2 \right> \right>  \to  2\eta_1 \left[ \eta_1 \Gamma + \eta_2 \left( \eta_2 \widetilde{T}_1 + \eta_1 \widetilde{T}_2 \right) \right] \left( 1 +  \Gamma \right),
\end{equation}
\begin{equation}
\frac{ 1 }{t_o} \left< \left< Q_2^2 \right> \right>  \to  2\eta_2 \left[ \eta_2 \Gamma + \eta_1 \left( \eta_2 \widetilde{T}_1 + \eta_1 \widetilde{T}_2 \right) \right] \left( 1 +  \Gamma \right),
\end{equation}
where the term proportional to $\Lambda B^2$ vanishes.
Indeed, we see that the external noise $F$ acts increasing the standard deviation of the distribution of $Q_i$. In particular, when $F$ is a compound Poisson noise, the fluctuations of $Q_i$ are affected by the rate of kicks and the statistics of noise intensities.
Thus, Gaussian or non-Gaussian stochastic forces influence in different ways the fluctuating behavior of thermal energies exchanged with heat baths.

\begin{figure}
 \centering
 \includegraphics[scale=0.5]{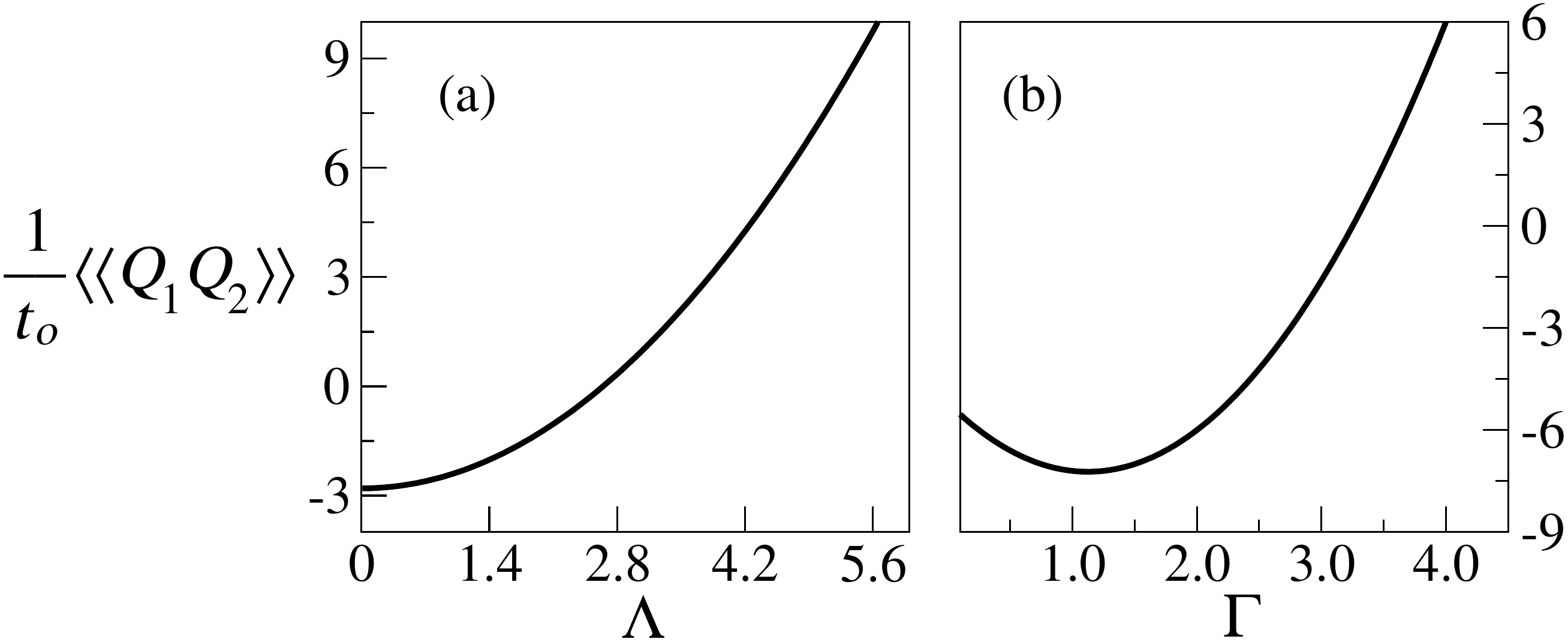}
 \caption{Effects of athermal noise on cumulant $\left<\left< Q_1 Q_2 \right>\right>$ over $t_o$. (a) Poisson noise with $\widetilde{T}_1 =2$, $\eta_1 = 0.2$, and $B = 0.5$. (b) Gaussian limit of external noise ($\Lambda \to \infty$, $B \to 0$, $\Lambda B = \Gamma$ fixed) with $\eta_1 = 0.5$, $\widetilde{T}_1=4$. The values of $\eta_2$ and $\widetilde{T}_2$ are obtained through \eqref{ScaledVar2}.}
\label{Cumuq1q2}
\end{figure}

We see that the covariance $\left<\left< Q_1 Q_2 \right>\right>$ can be positive or negative; the sign depends on the values taken by the parameters of the model, as shown in Fig. \ref{Cumuq1q2} (a).
Since the cumulant $\left<\left< Q_1 Q_2 \right>\right>$ may change its sign, it is possible that $\left<\left< Q_1 Q_2 \right>\right>$ can be zero: we may have uncorrelated heat fluctuations. However, even in such cases, $Q_i$ and $W$ are correlated, as long as $\eta_i$ is non-zero and the effects of external noise are not negligible.

In the absence of the athermal reservoir, which is equivalent to taking $\Lambda B=0$, the covariance of $Q_1$ and $Q_2$ is negative, 
\begin{equation}
 \frac{1}{t_o}\left<\left< Q_1 Q_2 \right>\right> = -2\eta_1\eta_2\left( \eta_2\widetilde{T}_1 + \eta_1\widetilde{T}_2 \right),
\end{equation}
and if the temperatures are different, heat flows on average from one thermal source to the other,
\begin{equation}
 \frac{1}{t_o}\left<\left< Q_1 \right>\right> = -\frac{1}{t_o}\left<\left< Q_2 \right>\right> = \eta_1\eta_2 \left( \widetilde{T}_1 - \widetilde{T}_2 \right).
\end{equation}
This is reasonable because $\left<\left< Q_1 Q_2\right>\right> = \left< \left( Q_1 - \left< Q_1 \right> \right)\left( Q_2 - \left< Q_2 \right> \right)\right>$ indicates how the fluctuations of $Q_1$ are correlated with the ones of $Q_2$: one thermal bath delivers and the other absorbs heat.
When the external noise is Gaussian, we have: 
\begin{equation} \label{CumuQ1Q2G}
\frac{1}{t_o} \left<\left< Q_1 Q_2 \right>\right> \to -2\eta_1\eta_2\left( \eta_2 \widetilde{T}_1 + \eta_1 \widetilde{T}_2 - \Gamma \right) \left( 1 + \Gamma \right),
\end{equation}
where $\Lambda \to \infty$, with $\Lambda B = \Gamma$ fixed. The quantity $\left<\left< Q_1 Q_2 \right>\right>$ is negative as thermal environments effects are more important ($\Gamma$ small); but it is positive when the effects of athermal reservoir dominate ($\Gamma$ large). Indeed, according to \eqref{Cumu1Qa}--\eqref{Cumu1Qb} with $\Gamma = \Lambda B$, the increasing of $\Gamma$ favors $\left< Q_i \right>$ negative. Then, both thermal reservoirs tend to receive heat due to the athermal energy injection when $\Gamma$ is large.

Also, observe that \eqref{CumuQ1Q2G} is a convex quadratic function of $\Gamma$, provided that the remaining parameters are fixed. Since $\Gamma$ is non-negative, one can show that \eqref{CumuQ1Q2G} is zero when $\Gamma_0 = \eta_2 \widetilde{T}_1 + \eta_1 \widetilde{T}_2 $, and there exists a minimum in the graph of $\left<\left< Q_1 Q_2 \right>\right>/t_o$ at $\Gamma_{min} = \left( \eta_2 \widetilde{T}_1 + \eta_1 \widetilde{T}_2 -1 \right)/2$, which is possible as long as $\left( \eta_2 -\eta_1 \right)\left( \widetilde{T}_1 - \widetilde{T}_2 \right)>0 $. Then, the covariance $\left<\left< Q_1 Q_2 \right>\right>/t_o$ may exhibit a non-monotonic behavior as a function of the parameter $\Gamma$ associated with the strength of the external Gaussian noise, see Fig. \ref{Cumuq1q2} (b). Nevertheless, it is important to bear in mind that the topography of the graph of $\left<\left< Q_1 Q_2 \right>\right>/t_o$ can be more complicated because of the dependence on other parameters, i.e., $\eta_i$ and $\widetilde{T}_i$. Notice these features are for $F$ being an external Gaussian noise.

\begin{figure}
 \centering
 \includegraphics[scale=1]{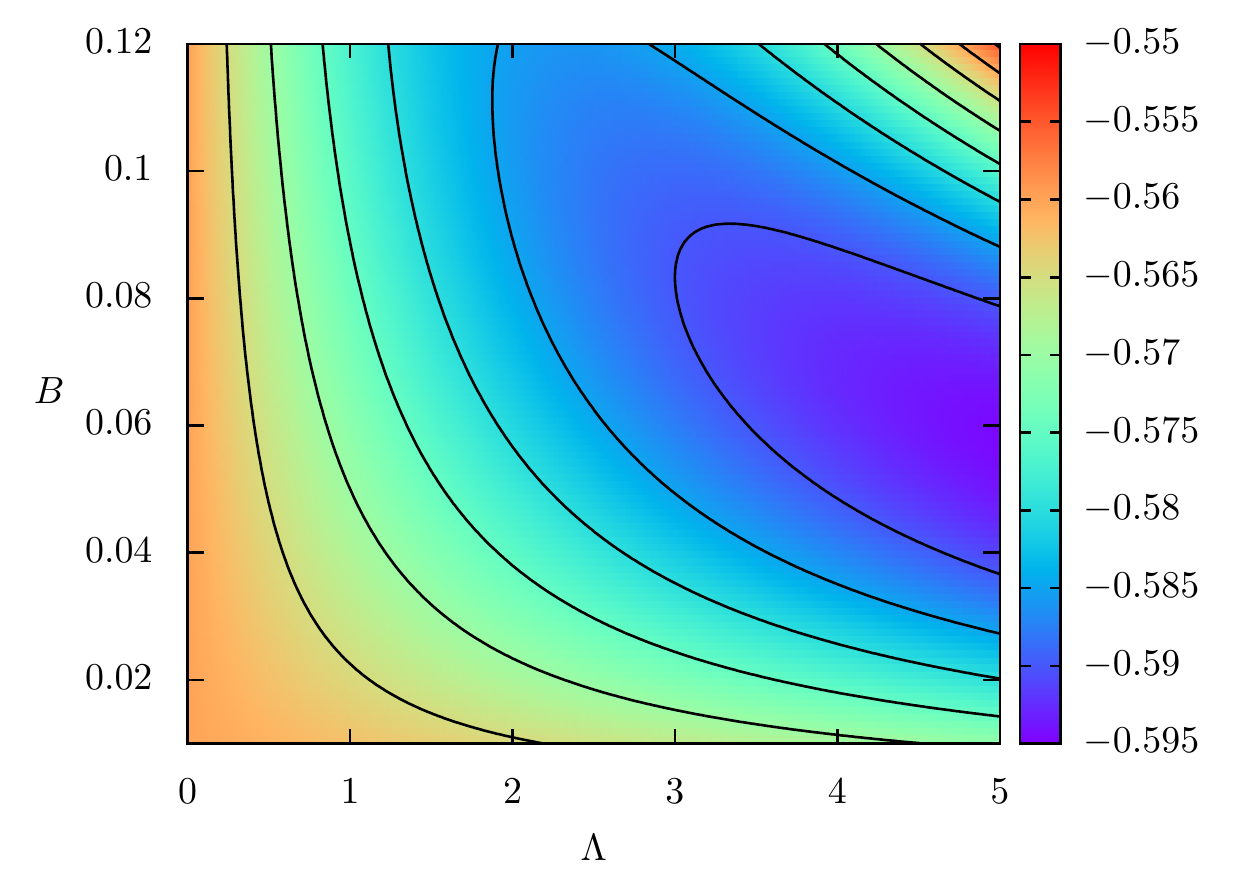}
 \caption{Contour plot of covariance $\left<\left< Q_1 Q_2 \right>\right>/t_o$ for $\widetilde{T}_1 =2$ and $\eta_1 = 0.2$. The values of $\eta_2$ and $\widetilde{T}_2$ are obtained through \eqref{ScaledVar2}.}
\label{CplotCumuq1q2}
\end{figure}

When $F$ is an athermal noise of Poisson kind, we perceive the emergence of a minimum in the graph of $\left<\left< Q_1 Q_2 \right>\right>$ in the region of $\Lambda$ large and $B$ small, as shown in Fig. \ref{CplotCumuq1q2}, which is consistent with the findings for the Gaussian limit of external noise. It is possible to interpret the sign of $\left<\left< Q_1 Q_2 \right>\right>$ by assuming that $\eta_i$ and $\widetilde{T}_i$ are fixed. Then, according to \eqref{CumuQ1Q2}, if $B$ is constant, i.e. for a given mean injected energy per Poisson impulse, we have that $\left< \left< Q_1 Q_2\right>\right>/t_o$ is negative when $\Lambda$ is small and positive for $\Lambda$ large. Thus, a very low density number of kicks $\Lambda$ with finite mean energy $B$ does not modify the long-run tendency of heat to flow from hot to cold thermal baths. On the other hand, by supposing a given Poisson rate $\Lambda$, we see that $\left< \left< Q_1 Q_2\right>\right>/t_o$ is negative if $B$ is small and positive when $B$ is large: the energetic effects of dominant athermal noise favor the absorption of heat by thermal baths.

We like to mention that our results are obtained by assuming $\Lambda$ and $B$ in \eqref{ScaledVar1} as independent quantities, and the Gaussian limit of external noise is obtained by taking $\Lambda \to \infty$, $B \to 0$, $\Lambda B$ constant. If we write $\Lambda B = \Gamma $ and consider that $\Lambda$ and $\Gamma$ are now the new independent non-dimensional parameters related to non-Gaussian noise effects, we would find apparently distinct results. For example, in the equations of high-order cumulants of $Q_i$ and $W$ shown in \eqref{CumuJ2}--\eqref{CumuJ3} and \eqref{Cumu2Q1}--\eqref{CumuQ1Q2}, we may rewrite $\Lambda B^n = \Gamma^n/\Lambda^{n-1}$, and the limit of small $\Lambda$ with $\Gamma$ fixed gives a quantity that is very large. We can deal with that by assuming $\Gamma$ also very small in order to find a $\Lambda B^n$ term that approaches zero.

Independent of the kind of noise adopted in our model system, the study of work and heat correlations should be done with care because we also have to consider the temperatures of heat baths. Indeed, our analysis shows that, for a two-temperature Langevin system, the inclusion of external noise may promote non-trivial fluctuations of energetic quantities. On the other hand, for a given non-deterministic system, the coupling to a single or many thermal baths can lead to interesting and unusual power fluctuations.

\section{Conclusion} \label{Con}

We investigated the non-equilibrium case of a damped Langevin-like model that represents an inertial Brownian particle in a harmonic potential and under the action of two thermal baths, at different temperatures, and an athermal reservoir, represented by an external Poisson noise. Since we have a linear stochastic model driven by independent noises, formal solutions are obtained without difficulties, which allow us to calculate all the time-dependent cumulants of position $X$ and velocity $V$ of the particle. With these cumulants, we also determined the cumulant-generating function of $X$ and $V$, as a function of time, as well as its expression for steady-states, which is given by an integral form calculated exactly.

We studied the stochastic properties of the energetic exchanges between the Brownian particle and the thermal and athermal reservoirs for stationary states. In fact, we determined the long-term behavior of first cumulants of the injected energy $W$ due to the external noise and the heat $Q_i$ related to thermal baths, which are defined by using appropriate dimensionless variables. The harmonically confined particle achieves an out-of-equilibrium steady-state with non-zero heat currents related to thermal baths and external power due to the athermal reservoir. 
For the injected athermal energy, which is a non-Gaussian variable, we shown that the standard deviation and the asymmetry (third cumulant) of the distribution of $W$ are influenced by the rate of Poisson kicks and the statistics of noise impulses, as well as the properties of thermal baths.
Also, we evaluated the covariance of heat exchanges, which present different behaviors if the external noise is Poissonian or Gaussian. This covariance, represented by the second cumulant $\left< \left< Q_1 Q_2\right> \right>$, can be positive, negative and even zero, in addition to have a non-monotonic behavior in terms of the parameters of the model. Thus, the specific structure of external noise plays a prominent role in the correlations and fluctuations of heat exchanges. 

\begin{acknowledgments}
This study was financed in part by the Coordenação de Aperfeiçoamento de Pessoal de Nível Superior - Brasil (CAPES) - Finance Code 001. 
\end{acknowledgments}

\bibliographystyle{apsrev4-1}
\bibliography{REF}

\end{document}